\documentclass{article}
\usepackage[utf8x]{inputenc}
\usepackage{spconf,amsmath,graphicx}

\graphicspath{{figures/}}

\usepackage{fancyhdr}
\usepackage{cite}
\usepackage{amsmath,amssymb,amsfonts}
\usepackage{algorithmic}
\usepackage{graphicx}
\usepackage{multirow}
\usepackage{tikz}
\usepackage{hyperref}
\usepackage{xcolor}
\usepackage{lipsum}
\usepackage{caption}
\usepackage{siunitx}
\usepackage{commath}
\usepackage{subfig}
\usepackage{pgfplots}
\pgfplotsset{compat=1.17}
\usepackage{mathrsfs}
\usetikzlibrary{arrows}
\usetikzlibrary{positioning}
\usetikzlibrary{3d}
\usepackage{comment}
\usepackage{adjustbox}
\usepackage{textcomp}
\usepackage{xcolor}
\def\BibTeX{{\rm B\kern-.05em{\sc i\kern-.025em b}\kern-.08em
    T\kern-.1667em\lower.7ex\hbox{E}\kern-.125emX}}
 
\fancypagestyle{firstpage}{
    \fancyhf{} 
    \fancyhead[L]{Submitted to IEEE ICIP 2024} 
}
 
\begin{document}

\thispagestyle{firstpage}

\definecolor{col_w}{rgb}{0.870588,0.796078,0.776470}
\definecolor{col_wn}{rgb}{0.996078,0.847059,0.364706}
\definecolor{col_out}{rgb}{1,0.5,0}
\definecolor{col_bc}{rgb}{0,0.5,1}
\definecolor{col_conv}{rgb}{0,1,1}
\definecolor{col_out}{rgb}{1,0.5,0}

\title{Motion-Adaptive Inference for Flexible Learned B-Frame Compression}

\name{M. Akin Yilmaz$^{\star}$ \qquad O. Ugur Ulas$^{\dagger}$ \qquad Ahmet Bilican$^{\dagger}$ \qquad A. Murat Tekalp$^{\dagger}$
\thanks{$^{\dagger}$This work is supported in part by TUBITAK 2247-A Award No.~120C156 and KUIS AI Center funded by Turkish Is Bank. A. M. Tekalp also acknowledges support from Turkish Academy of Sciences (TUBA).}}

\address{$^{\star}$ Codeway AI Research \\
      $^{\dagger}$ Dept. of Electrical \& Electronics Engineering, Koç University, Istanbul, Turkey 
}

\maketitle
\begin{abstract}
While the performance of recent learned intra and sequential video compression models exceed that of respective traditional codecs, the performance of learned B-frame compression models generally lag behind traditional B-frame coding. The performance gap is bigger for complex scenes with large motions. This is related to the fact that the distance~between the past and future references vary in hierarchical B-frame compression depending on the~level of hierarchy, which causes motion range to vary. The inability of a single B-frame compression model to adapt to various motion ranges causes loss of performance. As a remedy, we propose controlling the motion range for flow prediction during inference (to approximately match the~range of motions in the training data) by downsampling video frames adaptively according to amount of motion and level of hierarchy in order to compress all B-frames using a single flexible-rate model. We present state-of-the-art BD rate results to demonstrate the superiority of our proposed single-model motion-adaptive inference approach to all existing learned B-frame compression models.~\footnote{The models and instructions to reproduce our results will be released at 
\url{https://github.com/KUIS-AI-Tekalp-Research-Group/video-compression/tree/master/ICIP2024}.}.
\end{abstract}


\section{Introduction}
\label{intro}
The landscape of image and video compression is undergoing a transformative shift with the advent of deep learning. While significant strides have been made in the performance of deep learning-based video compression for  intra-frame and sequential (low-delay) video coding, learned hierarchical B-frame compression still poses a unique set of challenges. This paper endeavors to address these challenges with a particular emphasis on mitigating the inference-time data drift observed in learned optical flow prediction across distant frames.

Traditional video codecs have been endowed with tools to optimize mode selection for each individual coding unit, thereby achieving a remarkable degree of content adaptation. However, as we venture into the realm of hierarchical B-frame coding, we are faced with the problem of training a single model to handle varying motion ranges due to different temporal distance between reference frames at different levels of temporal hierarchy. Unlike sequential coding, where motion vectors for successive frames generally have uniform range, training a single model for hierarchical B-frame compression with a wide range of motion vectors causes loss of compression efficiency. This raises a fundamental question: Can learned hierarchical B-frame coding with a single model surpass the performance of traditional B-frame coding that excels in content adaptation on a frame-by-frame basis?

In this paper, we propose a basic motion adaptation strategy of adaptive downsampling of reference frames at inference to bring motion ranges to a scale observed in the training data and mitigate data drift during flow prediction stage. Through a comprehensive performance evaluation of the proposed motion adaptation strategy, we aim to bridge the performance gap between learned models and traditional codecs for hierarchical B-frame coding.
In doing so, this paper offers valuable insights for future research in the pursuit of content-adaptive and more efficient hierarchical B-frame coding.
\vspace{-4pt}

\section{Related work and Contributions}

\subsection{Neural Sequential (Low-Delay) Video Compression}

Early deep learning-based video compression models mainly focused on sequential (low-delay) coding by replacing all components of a traditional sequential video codec with jointly optimized subnetworks. A significant advance in this domain was made by Agustsson et al.~\cite{agustsson_scale}, who introduced a scale-space flow model for motion compensation that could account for motion uncertainty, including occlusions. A subsequent study~\cite{elfvc} proposed an innovative extension of the scale-space flow concept, integrating an in-loop flow predictor and a groundbreaking backbone architecture for analysis and synthesis transformations.

Ladune et al.~\cite{ladune2021conditional} introduced conditional coding, which was a radical shift from traditional residual coding, to obtain remarkable performance improvements. Conditional coding adopts a complex nonlinear function, to replace the simple subtraction used in conventional residual coding, for fusing the motion-compensated frame and the current frame for the best compression performance.
Li et al.~\cite{li2021deep} advanced this approach further by substituting motion-compensated frames with learned contextual features. Later, Sheng et al.~\cite{sheng2022temporal} introduced a temporal context mining module. This module is innovative in its approach to learning temporal contexts from propagated features, as opposed to relying on context generated from previously decoded frames. 

The significance of a powerful entropy model on coding efficiency performance cannot be overstated. Recently, a versatile entropy model was proposed by Li et al.~\cite{li2022hybrid}, which effectively captures both temporal and spatial correlations, resulting in enhanced Rate-Distortion (RD) performance.

More recently, Li et al.~\cite{li2023neural} introduced the concept of offset diversity combined with cross-group interaction. This concept is particularly effective in addressing complex motion alignments and enriching diversity in the temporal dimension. Complementing these temporal advances, our work proposes a finely detailed quadtree-based partitioning method. This method significantly enhances spatial context diversity, marking a notable contribution to the field.

\subsection{Neural Bi-Directional Video Compression}
Research into learned bi-directional (random access) video compression remains in its early stages, despite the well-known fact that traditional hierarchical B-frame coding yields superior RD performance over sequential coding. One of the early works~\cite{hlvc} employs a tri-layered hierarchical quality system complemented by a recurrent neural network for post-processing to enhance the visual fidelity of the output.

In concurrent research endeavors~\cite{ours_icip20, lhbdc}, the authors have made significant strides by implementing in-loop bi-directional flow prediction. This technique is paired with a sophisticated learned fusion mask, which adeptly blends both forward and backward motion-compensated frames. The result is a remarkably smoother residual frame that lends itself more readily to compression.

Building upon these concepts, \cite{ladune2021conditional} puts forth a versatile model capable of processing I, P, and B frames with equal aplomb, utilizing conditional coding. In\cite{flexrate}, the replacement of pre-trained flow estimation models aims a more direct estimation of motion residuals. This is achieved through a compression bottleneck analogous to that in~\cite{agustsson_scale}, with the addition of a nuanced frame-level rate control mechanism. The most recent contributions to this field, detailed in~\cite{ours_icip23}, involve the implicit motion compensation via deformable convolutions\cite{deformv2} at multi-scale feature maps utilizing a single bottleneck for both predicting and compressing deformable offsets and feature level residuals.  Nowadays, implicit neural representation models like \cite{hinerv} have also shown a comparable performance with state-of-the-art (SOTA) learned codecs even though it is not end-to-end optimized.

\begin{figure*}[ht]
\centering
 \includegraphics[width=1\textwidth]{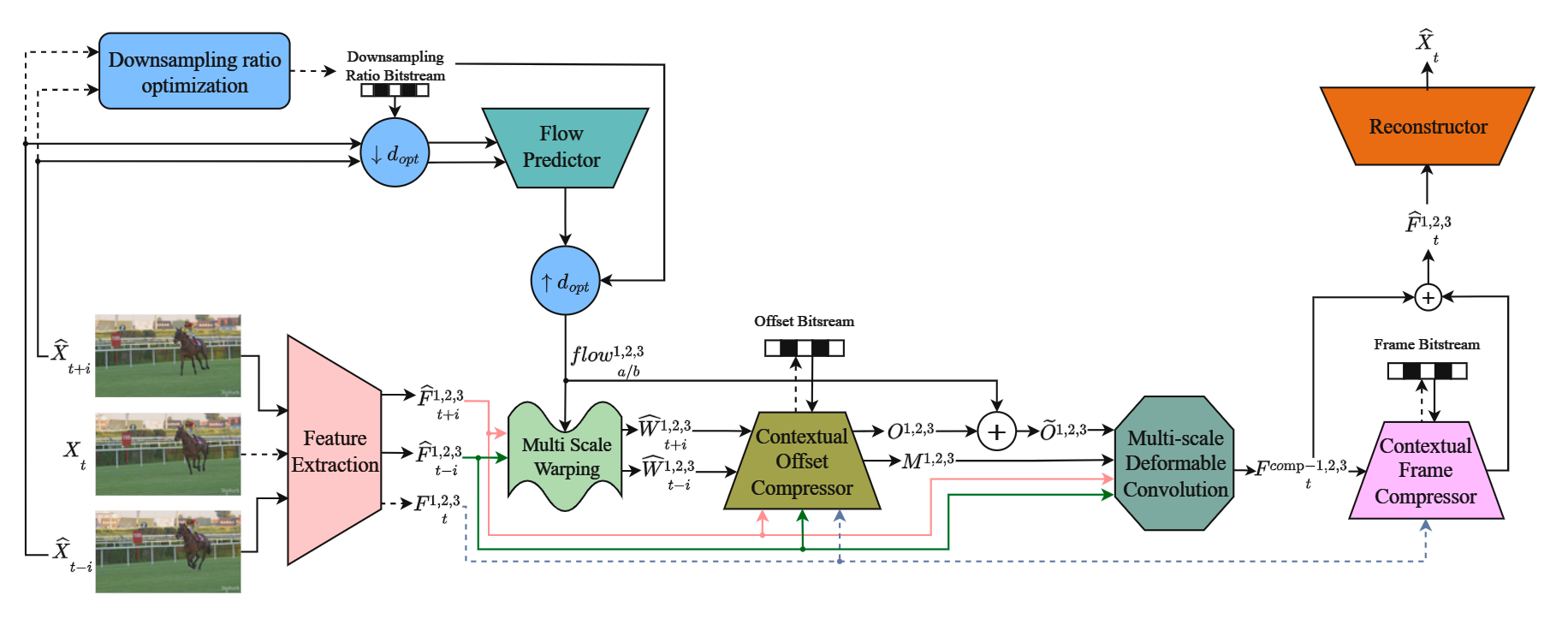} 
 \vspace{-25pt}
\caption{Overview of the proposed motion-adaptive bi-directional (B-frame) compression model. $t$ and $i$ denote the current frame index and past/future reference frame distance, respectively. e.g., for GoP size 16, $i$ can be one of (1, 2, 4, 8). Superscripts~\{1,2,3\} denote the scale level. Decoded variables are represented by $\hat{\cdot}$. Dotted lines are present only in the encoder.}
\label{fig:arch}
\end{figure*}

\subsection{Flow-Guided Deformable Convolutions}
Deformable convolution has emerged as a powerful tool to handle geometric transformations between video frames, but it suffers from training instability due to offset overflow. To address this, BasicVSR++ \cite{chan2022basicvsrpp} integrates optical flow to guide the Deformable Convolutional Network (DCN), enhancing stability. It was demonstrated that this approach surpasses traditional flow-based frame alignment by leveraging the diversity of offsets. Two advantages are clear: firstly, the pre-alignment of features via optical flow facilitates the learning of offsets within CNN's local receptive fields. Secondly, focusing on residual learning allows the network to concentrate on small deviations, easing the burden on the deformable alignment module.

\subsection{Content Adaptation}
Current learning-based codecs often struggle with adaptability to different video contents, creating a disparity between training and testing datasets. This issue is addressed through methods like online encoder update strategies, as seen in \cite{ours_icip23,guo_content_adaptive} allowing dynamic adjustment of encoder parameters based on content, without complicating the decoder. Additionally, the challenge of frame reference structure in video coding is tackled by adopting bit allocation strategies. This involves determining optimal $\lambda$ values for different video frames or regions, enhancing compression efficiency by focusing resources on more complex or important areas. While some methods use a constant lambda offset \cite{li2023neural,flexrate}, more sophisticated approaches, like those in \cite{bit_allocation_2023} employ optimization techniques for more precise $\lambda$ adjustments. These advancements highlight the ongoing need for content adaptation, dynamic parameter adjustment, and strategic bit allocation in video compression.

\vspace{-6pt}

\subsection{Contributions}
This paper extends our earlier research in hierarchical B-frame compression~\cite{ours_icip20,ours_icip23} to obtain state-of-the-art results. Specific key contributions in this paper include:

1) Motion-adaptive flow prediction to handle data drift due to varying motion range: Efficient hierarchical B-frame compression relies on effective flow prediction between the past/future references and the current frame. Our proposed adaptive frame downsampling strategy brings the range of motion for different videos and frames with different distances to the current frame to the scale observed in the training set and improves accuracy of predicted flows.

2) Prediction driven flow-guided multi-scale off-set estimation for deformable convolutions: We perform flow-guided offset estimation (similar to that proposed in \cite{chan2022basicvsrpp}) in the new use case of deformable convolution based current frame feature prediction, by using predicted flows instead of actual flows since the current frame is not available at the~decoder.

\pagebreak

\section{METHOD}
\vspace{-3pt}
\subsection{Overview of Model Architecture} \label{framework} \vspace{-3pt}
The proposed model aims to compress an intermediate frame denoted as $X_{t}$ using a past decoded reference frame $\hat{X}_{t-i}$ and a future decoded reference frame $\hat{X}_{t+i}$. The overall flow diagram for the model is shown in Figure\ref{fig:arch}, which extends our previous work~\cite{ours_icip23} to perform motion-adaptive inference. The proposed model contains several newly introduced and modified sub-networks, which are briefly summarized here. They are explained in more detail in Sections~\ref{adaptive_motion_pred}-\ref{contextcoding}.

The {\bf Feature Extractor} extracts multi-scale feature maps, $\hat{F}^{1,2,3}_{t+i}, \hat{F}^{1,2,3}_{t-i}, F^{1,2,3}_{t}$, for decoded future reference $t+i$, decoded past reference $t-i$ and the current frame $t$, respectively, independently by applying multiple strided convolutions and residual blocks. The superscripts denote the scale level.

The {\bf Motion-Adaptive Flow Predictor} is a UNet that generates estimates of flow vectors between the past and future references in both directions. Then, the forward flow estimate from the past reference to the current frame is set to half of the flow from the past to the future reference, likewise the backward flow estimate from the future reference to the current frame is set to half of that from the future to the past reference.
Since we employ adaptive frame down-sampling, as explained in Section~\ref{adaptive_motion_pred} in detail, we estimate bi-directional flows at the same resolution as the input frames. This step relies solely on decoded reference frames; hence, there is no bitrate overhead to send flow data. Denoting estimated flows as $flow^{0}_{a/b}$, we down-sample it by 2, 4, 8 to obtain $flow^{1,2,3}_{a/b}$, respectively, aligned with the resolutions of the multi-scale feature maps. We then apply multi-scale feature warping, warping feature maps of both references towards the current frame using bidirectional flow estimates.

The {\bf Contextual Offset Compressor} is a compressive autoencoder. The encoder sub-net takes the feature maps of both reference frames and their warped feature maps $\hat{W}^{1,2,3}_{t-i}, \hat{W}^{1,2,3}_{t+i}$, as well as the feature map for the current frame as inputs and generates a bitstream. The decoder sub-net takes these bits as well as the available condition data at the decoder side and outputs multi-scale offsets $O^{1,2,3}$ and modulation scalars $M^{1,2,3}$. Subsequently, these multi-scale offsets undergo refinement guided by multi-scale flow estimates. More details can be found in Section~\ref{offsetcoding}.

Three separate deformable convolution blocks, each operating at a different resolutions, process both reference feature maps using the refined offsets and modulation scalars. This process yields the predicted feature maps for the current frame, denoted by $F^{comp-1,2,3}_{t}$.

The {\bf Contextual Frame Compressor} is the second compressive autoencoder. Its encoder sub-net takes $F^{comp-1,2,3}_{t}$ and $F^{1,2,3}_{t}$ as inputs, and generates a bitstream. Its decoder sub-net takes these bits in addition to $F^{comp-1,2,3}_{t}$ as input and outputs a decoded feature map for the current frame. Details are provided in Section~\ref{contextcoding}.

This reconstructed feature map is processed by multiple blocks, including a pixel shuffler layer and residual blocks, ultimately producing the final decoded current frame $\hat{x}_{t}$.
\vspace{-6pt}

\subsection{Motion-Adaptive Flow Prediction} \label{adaptive_motion_pred} \vspace{-3pt}
The performance of deep learning-based video compression models inherently relies on learned distributions from the~training set. While models excel when inference data is within the range of the training data, their performance falters when extrapolating to out-of-disribution data. This~issue becomes particularly pronounced in bidirectional video compression, where the model's effectiveness diminishes with expanded intra period or when handling videos with substantially higher motion than the training videos. 

To mitigate the effect of data drift in the Flow Predictor due to varying motion vector ranges, we propose motion-adaptive flow prediction. "Motion-adaptative inference" refers to adaptively selecting the resolution scale that flow prediction is performed at the encoder. By downsampling both the past and future reference frames, the range of motion vectors between them can be controlled, effectively aligning the distribution of flow vectors during inference with the~distribution learned during training.
\vspace{3pt}

\noindent \textbf{Optimization of Down-sampling Factor. } 
The Flow Predictor operates both in the encoder and decoder. The encoder selects the down-sampling factor that best mitigates data drift out of the list $\left\{ 1,2,4,8 \right\}$. The best downsampling factor for each coded frame is signalled to the decoder using 2 bits.

We observed that the quality of the prediction made with predicted flow is directly proportional to compression performance. 
Hence, we choose the best out of the four different down-sampling factors $d\in \left\{ 1,2,4,8 \right\}$ using the following procedure: \vspace{-3pt}
\begin{enumerate}
\item Estimate flow $flow_{a/b}(d)$ by first downsampling both reference frames by a factor $d$ and then upsampling and adjusting the magnitude of the calculated motion \vspace{-3pt}
\begin{equation}
flow_{a/b}^0(d)  = \uparrow_d (FlowPredictor(\downarrow_d \hat{X}_{t+i}, \downarrow_d \hat{X}_{t-i})) \nonumber  \vspace{-3pt}
\end{equation}  
where \( flow^0_a \) denotes flow from \( \hat{X}_{t-i} \) to \( \hat{X}_{t+i} \), and \( flow^0_b \) is the flow from \( \hat{X}_{t+i} \) to \( \hat{X}_{t-i} \).
\item Compute the warped images $\overline{X}_{t-i}$ and $\overline{X}_{t+i}$\vspace{-3pt}
\begin{equation}
\begin{split}
    & \overline{X}_{t+i} = Warp(\hat{X}_{t+i}, 0.5flow^0_a) \\
    & \overline{X}_{t-i} = Warp(\hat{X}_{t-i}, 0.5flow^0_b) \\
\end{split}
\end{equation} 
\vspace{-14pt}
\item Compute prediction PSNR
\vspace{-4pt}
\begin{equation}
PSNR(d) = psnr(X_t,0.5(\overline{X}_{t+i}+\overline{X}_{t-i}))   \nonumber
\end{equation} \vspace{-22pt}
\item Choose the factor $d_{opt}$ that yields the best PSNR \vspace{-3pt}
\begin{equation}
d_{opt} = argmax(PSNR(d))
\end{equation} 
\end{enumerate}
\vspace{-6pt}

\begin{figure}[t]
\centering
\begin{tabular}{cc}
  \includegraphics[width=0.218 \textwidth]{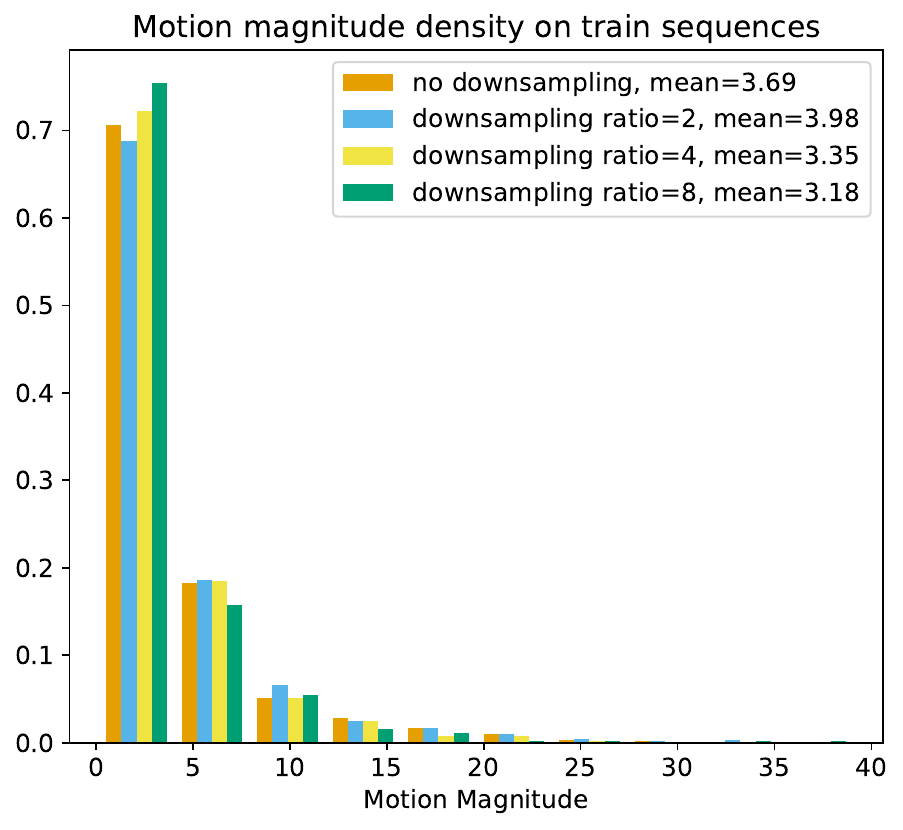} &
  \includegraphics[width=0.218\textwidth]{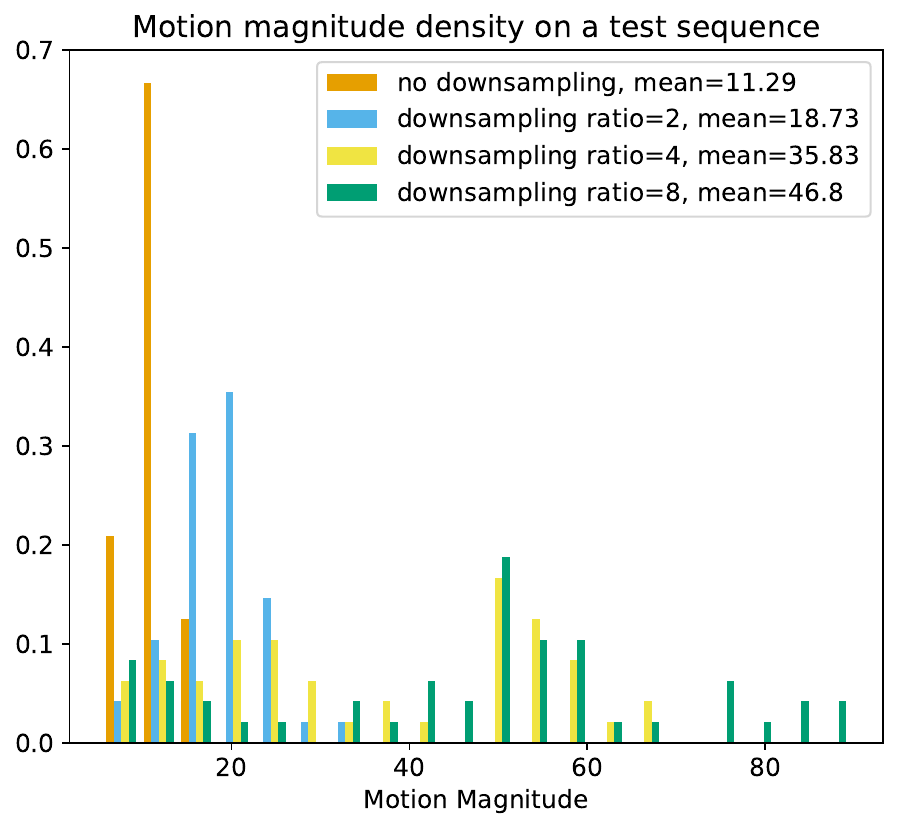} 
\end{tabular}
\caption{The mean over all frames of flow-magnitude histograms predicted by the \textit{Flow Predictor} for the train set (left) and a test sequence with large motion (right).}
\label{fig:downsampling}
\vspace{-8pt}
\end{figure}

By downsampling the reference frames before flow prediction during inference and then upscaling the estimated flows accordingly, our model gains the~ability to predict flows with magnitude greater than those seen in the training~set. The success of the proposed motion-adaptive flow prediction is demonstrated in Figure~\ref{fig:downsampling}. On the left, we see the range of predicted flows in the training set is relatively small. On the right, we see that the proposed adaptive Flow Predictor can predict larger flows in the test sequence relatively well by the virtue of adaptive downsampling during inference. This is because the motion in the frames downsampled by 4 aligns well with the range of motion encountered in the training set, where it achieves good predictions.

\noindent \textbf{Multi-scale Warping. }
We apply linear warping to warp feature maps of reference frames towards the current frame~as:
\begin{equation}
\begin{split}
    & \hat{W}^j_{t+i} = Warp(\hat{F}^{j}_{t+i}, 0.5flow^j_a) \\
    & \hat{W}^j_{t-i} = Warp(\hat{F}^{j}_{t-i}, 0.5flow^j_b) \\
\end{split}
\end{equation}
The superscripts \( j \in \{1, 2, 3\} \) denote the scale level of feature maps and flow fields.

The relation between the \( flow^j_{a/b} \)'s is given by:
\begin{equation}
    flow^{j+1}_{a/b} = 0.5(\downarrow 2(flow^j_{a/b}))
\end{equation}
where $\downarrow$2 represents bilinear downsampling.  
The warped maps \( \hat{W}^j_{t-i} \), \( \hat{W}^j_{t+i} \), along with \( \hat{F}^{1,2,3}_t \), \( \hat{F}^{1,2,3}_{t-i} \), and \( F^{1,2,3}_{t+i} \)  constitute the context for the contextual offset compressor. 
\vspace{3pt}

\subsection{Prediction-driven Flow-guided Multi-scale Offset Estimation and Compression}
\label{offsetcoding}
While motion-compensation in feature space using deformable convolutions has been used in sequential video compression~\cite{FVC2022} and use of optical flow to guide offsets in deformable convolutions to enhance training stability was proposed in BasicVSR++~\cite{chan2022basicvsrpp}, we combine these two concepts in the context of hierarchical bi-directional frame compression. 
Unlike BasicVSR++, in video compression, the current frame is not available at the decoder. To remedy this, we show that replacing actual flows with predicted flows is equally effective to guide offsets in deformable feature compensation.

\noindent \textbf{Multi-scale Conditional Offset Compression. } A bottleneck autoencoder is used to estimate/compress the offsets $O^{1,2,3}$ and masks $M^{1,2,3}$ for each scale level by conditioning on both the warped and original feature maps at each level. 

A notable challenge is the tendency of offsets to learn the~noise in the data, often leading to performance degradation. To mitigate this, we apply \( \tanh(\cdot) \) function to the offsets, effectively curbing this adverse phenomenon. After contextual offset compression and decompression, we guide the offsets with flows for the deformable convolution given by:
\begin{equation}
    \tilde{O}^j = flow^j_{a/b} + \alpha^j tanh(O^j)
\end{equation}
where $\alpha^j$ is a scalar for scale level $j$.
Then, the masks $M^{1,2,3}$ and the flow-guided offsets $\tilde{O}^{1,2,3}$ are input into the multi-scale deformable convolutional network
\begin{equation}
    F^{comp-1,2,3} = DCN(M^{1,2,3},\tilde{O}^{1,2,3}) \label{Fcomp}
\end{equation}
to obtain compensated feature map for the current frame.

\subsection{Multi-scale Contextual Coding by Feature Prediction using Deformable Convolutions} 
\label{contextcoding}
We adopt multi-scale conditional coding, similar to our previous work~\cite{ours_icip23}, due to its proficiency in deriving efficient representations of video frames. This technique significantly enhances the quality of reconstructed videos, offering superior adaptability to the diverse characteristics inherent in different video content. Furthermore, conditional coding optimizes bitrate allocation by astutely discerning the importance of various regions within a frame. 

The encoding and decoding process involves conditional coding of multi-scale feature maps \( F^{1,2,3}_t \) for the current frame using \( F^{comp-1,2,3}_t \) as a condition through a single bottleneck autoencoder subnet $Enc$
\begin{equation}
    \hat{y}_t = round(Enc(F^{1,2,3}_t, F^{comp-1,2,3}_t))
\end{equation}
where \( F^{comp-1,2,3}_t \) denotes the outputs from the deformable convolutional network~(\ref{Fcomp}).
Subsequently, current frame features are decoded using the~decoder subnet $Dec$ with inputs~\( \hat{y}_t \) and  condition \( F^{comp-1,2,3}_t \) and reconstructed as  
\begin{equation}
\hat{F}_t^{1,2,3} = Dec(\hat{y}_t, F^{comp-1,2,3}_t) + F^{comp-1,2,3}_t 
\end{equation}
The reconstructor $Rec$ then yields the decoded video frame
\begin{equation}
 \hat{X}_{t} = Rec(\hat{F}_t^{1,2,3}) 
\end{equation}

\begin{table*}[]
\centering
\caption{BDBR(\%) performances of different learned codecs and our proposed method over UVG sequences. The anchor is the reference codec of H.266\cite{vtm}: VTM - 18.0 with random access and intra period = 16 configuration.} \vspace{-5pt}
\begin{adjustbox}{width=0.74\textwidth}
\begin{tabular}{|l|c|c|c|c|c|c|}
\hline
 &
  Ours &
  \begin{tabular}[c]{@{}c@{}}Li\\ 2023\cite{li2023neural}\end{tabular} &
  \begin{tabular}[c]{@{}c@{}}Kwan\\ 2023\cite{hinerv}\end{tabular} &
  \begin{tabular}[c]{@{}c@{}}Yilmaz\\ 2023\cite{ours_icip23}\end{tabular} &
  \begin{tabular}[c]{@{}c@{}}Li\\ 2022\cite{li2022hybrid}\end{tabular} &
  \begin{tabular}[c]{@{}c@{}}Sheng\\ 2021\cite{sheng2022temporal}\end{tabular} \\ \hline
Beauty           & -56.12         & -53.67         & -31.31         & -52.45         & -51.21        & -39.01          \\ \hline
Bosphorus        & -8.70          & 20.83          & 0.91           & 0.70           & 40.86         & 78.18          \\ \hline
Honeybee         & -7.94          & 20.56          & -76.03         & -7.89          & 33.47         & 63.84          \\ \hline
Jockey           & 46.11          & 59.88          & 90.43          & 114.92         & 71.86         & 171.07          \\ \hline
ReadySetGo       & 19.00          & 2.05           & 40.36          & 41.55          & 12.37         & 46.97          \\ \hline
ShakeNDry        & -7.71          & -2.70          & -30.06         & -4.03          & 7.71          & 27.16          \\ \hline
YachtRide        & -13.33         & -10.11         & 53.81          & -11.10         & -0.15         & 20.75          \\ \hline
\textbf{Average} & \textbf{-4.10} & \textbf{5.26} & \textbf{6.87} & \textbf{11.67} & \textbf{16.42} & \textbf{52.71} \\ \hline
\end{tabular}
\end{adjustbox}
\end{table*}


\section{Experimental Results}
\subsection{Experimental Setup} 
\textbf{Datasets.} We used Vimeo-90k dataset\cite{vimeo} to train our model similar to the literature. Each sequence in the dataset contains 7 consecutive frames with a resolution of  $448 \times 256$. We benchmarked our model performance on the UVG\cite{uvg} dataset, which is also commonly used in the literature. The~resolution of test sequences is $1920 \times 1080$. \\
\textbf{Training Details.} The training process involves a nuanced approach to frame selection, data augmentation, and rate distortion loss optimization. Frames are selectively chosen at regular intervals, adopting a 3-frame configuration from a sequence of 7 frames (1-3-5, 2-4-6, 3-5-7) to capture substantial motion variations within the video sequence. Temporal flipping is employed as a data augmentation strategy to enhance the model's robustness by introducing variations in temporal orientation and ensuring effective handling of diverse video content. The model is trained on $256 \times 256$ random crops with an end-to-end fashion using the Adam\cite{adam} optimizer and a rate distortion loss: $\lambda D + R$ where $D$ is calculated as the MSE between the ground truth and reconstructed frame and $R$ is the total rate of all compressive bottlenecks. In the initial 500,000 steps, the learning rate is set to 1e-4, with a primary focus on training the model to compress the middle frame while utilizing the other two frames as reference frames. Subsequently, in the next 500,000 steps, the learning rate is reduced to 1e-5. Post the initial training phase, the model undergoes additional training on 5 consecutive frames, utilizing the same data augmentation technique. The decoding process involves using the initially decoded frame as a reference, enhancing the model's ability to learn optimal rate allocation within group-of-pictures (GOP) structures and contributing to improved compression efficiency. The rate distortion loss is computed as the average of losses calculated for the 3-frame configuration, serving as the basis for optimization. This detailed training methodology provides insights into the effective integration of temporal flipping data augmentation in deep learning-based video compression, demonstrating the potential for achieving efficient compression with enhanced rate allocation within GOP structures. To enable variable rate compression within a single model, we follow the method explained in\cite{ours_icip23}. With this approach, we can interpolate between distinct points during the inference, and can achieve a varying compression ratio. For intra frame compression, we employed the same learned compression setup in \cite{ours_icip23}.

\subsection{Comparison with the State-of-the-Art}
We evaluate our model in terms of peak signal-to-noise ratio (PSNR) versus bits-per-pixel (bpp). We report PSNR and bpp results averaged over each decoded frame~\cite{psnr_comp} following the literature. To ensure an equitable comparison, Group of Pictures (GoP) sizes are set to 16 across all models.  

The BD-BR (Bjøntegaard Delta Bit Rate) results for the UVG sequences are tabulated in Table 1. Our model demonstrates a significant reduction in BD-BR, achieving a $-9.36\%$ reduction compared to the next best-performing learned codec. Also in comparison to VTM-18.0, our model achieves a $-4.10\%$ BD-BR reduction, underscoring its superior efficiency in video compression. Notably, it stands out as not only the best among learned codecs but also the sole learned codec outperforming VTM-18.0. 

\begin{table}[!t]
\centering
\caption{BDBR(\%) performances of our proposed method with and without motion adaptive downsampling and flow prediction. The anchor is VTM - 18.0 with random access and intra period = 16 configuration. AD, FP represents \textit{adaptive downsampling} and \textit{flow prediction} respectively}
\begin{adjustbox}{width=0.44\textwidth}
\begin{tabular}{|l|c|c|c|}
\hline
 &
  \begin{tabular}[c]{@{}c@{}}Ours /w\\ AD\end{tabular} &
  \begin{tabular}[c]{@{}c@{}}Ours /wo\\ AD\end{tabular} &
  \begin{tabular}[c]{@{}c@{}}Ours /wo\\ FP\end{tabular} \\ \hline
Beauty           & -56.12         & -56.13        & -55.68         \\ \hline
Bosphorus        & -8.70          & -7.84         & -1.38          \\ \hline
Honeybee         & -7.94          & -7.92         & -4.95          \\ \hline
Jockey           & 46.11          & 84.42         & 105.66         \\ \hline
ReadySetGo       & 19.00          & 40.57         & 71.94          \\ \hline
ShakeNDry        & -7.71          & -7.80         & -8.63          \\ \hline
YachtRide        & -13.33         & -12.44        & -9.27          \\ \hline
\textbf{Average} & \textbf{-4.10} & \textbf{4.69} & \textbf{13.96} \\ \hline
\end{tabular}
\end{adjustbox}
\label{table:ablation}
\end{table}

\subsection{Ablation Studies}

In Table~\ref{table:ablation}, we present the results of our comprehensive investigation into the effects of various components within the proposed method for video compression. Our findings indicate that incorporating flow prediction significantly enhances compression efficiency, particularly in sequences characterized by predictable, fast and abundant motion. The impact of flow prediction is most pronounced in sequences such as \textit{ReadySetGo} and \textit{Jockey}, which exhibit reductions in BD-BR of $31.37\%$ and $21.24\%$, respectively. Conversely, sequences like \textit{ShakeNDry}, and \textit{Beauty}; which lack camera motion and feature unpredictable movement coupled with less motion, show a negligible improvement with flow prediction, increasing the BD-BR rate by $0.83\%$ and decreasing it by $0.45$, respectively. However, in scenarios having slow motion with global motion exemplified by the \textit{Bosphorus} sequence a BD-BR reduction of $6.46\%$ is observed. On average, implementing flow prediction has reduced the BD-BR by $9.27\%$.

The application of motion adaptive reference frame downsampling, described in Section~\ref{adaptive_motion_pred}, further enhances compression performance, especially in sequences with complex and substantial motion. This technique yields a BD-BR reduction of $38.31\%$ in \textit{Jockey} and $21.57\%$ in \textit{ReadySetGo}, demonstrating its effectiveness in sequences with fast motion. The overall impact of adaptive downsampling across all sequences has led to an average BD-BR reduction of $8.79\%$.

In our sequence-by-sequence analysis, distinct characteristics and performances emerge across various test sequences. In the \textit{Beauty} sequence, the prowess of learned codecs becomes evident, outperforming traditional codecs mainly because it has simple motion. The \textit{Bosphorus} and \textit{YachtRide} sequences showcase the superiority of bidirectional video compression due to making use of future reference frames. For \textit{Honeybee} and \textit{ShakeNDry}, the work in\cite{hinerv} achieves optimal results, particularly because these sequences exhibit high frame similarities, ideally suiting implicit video modeling techniques. In \textit{Jockey,} our method excels among learned codecs. The bidirectional approach provides ample information for current frame modeling, yet its efficacy is somewhat constrained by the training dataset, leading to data drift. This issue is partially mitigated through motion adaptive downsampling, although traditional bidirectional compression maintains an edge under these conditions. In the \textit{ReadySetGo} sequence, containing significant occlusion, absent in our Vimeo90k training dataset, hinders the performance of learned bidirectional codecs. These codecs struggle when the distance between reference frames are large, resulting in sequential coding yielding more robust results. 

\section{Conclusion}
\label{conc}
We propose the first learned bi-directional  (B-frame) compression model that outperforms the random access mode of the H.266 reference codec VTM-18.0, consequently,  outperforming all available learned bi-directional coding models, over the UVG test set. The proposed motion adaptive frame downsampling for bi-directional flow prediction is the pivotal factor in the success of our framework, enabling effective modeling of complex video dynamics and addressing data drift between training and real-world test conditions.

Looking ahead, there is still room for further improvement, especially for sequences with localized high-motion such as Jockey and ReadySetGo. We continue to explore methods for content-specific adaptability at inference time to achieve  better generalization across diverse content types in challenging conditions.

\clearpage
\bibliography{references}
\bibliographystyle{IEEEtran}

\end{document}